\documentclass[10pt]{article}
\usepackage{makeidx}
\usepackage{multirow}
\usepackage{multicol}
\usepackage[dvipsnames,svgnames,table]{xcolor}
\usepackage{graphicx}
\usepackage{ulem}
\usepackage{amsmath}
\usepackage{amssymb}
\author{Noah}
\title{}
\usepackage[paperwidth=612pt,paperheight=792pt,top=72pt,right=72pt,bottom=72pt,left=72pt]{geometry}

\makeatletter
	{\par\setlength{\parindent}{#3}
	\setlength{\leftmargin}{#1}       \setlength{\rightmargin}{#2}%
	\advance\linewidth -\leftmargin       \advance\linewidth -\rightmargin%
	\advance\@totalleftmargin\leftmargin  \@setpar{{\@@par}}%
	\parshape 1\@totalleftmargin \linewidth\ignorespaces}{\par}%
\makeatother

\begin{document}

\title{Generalization of the PageRank Model}
\author{Noah E. Friedkin\textsuperscript{1}} 
\maketitle

\noindent \textit{ \textsuperscript{1} Center for Control, Dynamical Systems and Computation, College of Engineering, and  Department of Sociology, College of Letters and Science, University of California, Santa Barbara  }.

\begin{abstract} 
\noindent This paper develops a generalization of the PageRank model of page centralities in the global webgraph of hyperlinks. The webgraph of adjacencies is generalized to a valued directed graph, and the scalar dampening coefficient for walks through the graph is relaxed to allow for heterogeneous values.  A visitation count approach may be employed to apply the more general model, based on the number of visits to a page and the page's proportionate allocations of these visits to other nodes of the webgraph. 
\end{abstract}

\section{Introduction}

In a multi-agent network, the total influences of the agents are related to the number and length of the walks in the network. A model of this relationship \cite{FJ1990} is 

\begin{align}
{\bf{V}} & = ({\bf{I}} + \alpha{\bf{W}} + {{\alpha}^2}{\bf{W}}^{2} + {{\alpha}^3}{\bf{W}}^{3} + ... )(1- \alpha) \notag \\
& = ({\bf{I}} - \alpha{\bf{W}})^{-1} (1- \alpha)
\end{align}

\noindent where ${\bf{W}}_{n \times n}$ is row-stochastic and  $0<\alpha<1$ is scalar. The $v_{ij}$ of ${\bf{V}}=[v_{ij}]$ correspond to the relative net influence of agent $j$ on agent $i$. Thus,

\begin{equation} 
{\bf{r}} = \frac{1}{n}{{\bf{V}}^{T}}{\bf{1}} , \;\; 
{{\bf{r}}^{T}} {\bf{1}} = 1, 
\label{eq:TEC}
\end{equation}

\noindent is the average relative net influence of agent $j$ on all agents of the system, and may taken as a measure of the centrality of node $j$ in the system  \cite[pp.1485 -1487]{Friedkin1991}. The average may be based on the $n$ values of each column, or $n-1$ values when the main diagonal values of ${\bf{V}}$ are excluded. If the model is expressed as follows \cite{Brin1998} 

\begin{equation} 
{\bf{r}}  = \frac{1-\alpha}{n} + \alpha{\bf{W}}^{T} {\bf{r}} 
\label{eq:PageRank}
\end{equation}

\noindent then equation \ref{eq:TEC} is its solution. 

\section{Generalization}

The generalization \cite{Friedkin1998} of equation \ref{eq:TEC} is
\begin{align}
{\bf{V}} & = ({\bf{I}} + {\bf{AW}} + ({\bf{AW}})^{2} + ({\bf{AW}})^{3} + ... )({\bf{I}} - {\bf{A}}) \notag \\
& = ({\bf{I}} - {\bf{AW}})^{-1} ({\bf{I}} - {\bf{A}})
\end{align}

\noindent where ${\bf{A}}$ is diagonal with each value $0<a_{ii}<1$. Clearly, this model includes the special case ${\bf{A}} = \alpha{\bf{I}}$, and its corresponding centrality values are also given by equation \ref{eq:TEC}. 

An agent with $a_{ii}$ near 0 or  $w_{ii}$ near 1 is in a receiver  (sink) state with outflows near zero. An agent with $a_{ii}$ near 1 and $w_{ii}$ near 0 is in a transmitter (flow-through) state. A useful implementation of the model couples $a_{ii}$ and $w_{ii}$. The simplest coupling is $a_{ii} = 1- w_{ii}$. 

With this coupling assumption, an implementation of the model may be based on the proportion of visits to a page $i$ that terminate on it, $w_{ii}$, and the proportions of visits to a page $i$ that activate a hyperlink to page $j \ne i$, $w_{ij}$. If such a specified  ${\bf{W}}$ is in hand, then ${\bf{A}}$ is empirically determined and its values reflect the extent to which a page is a sink or transmitter. Thus, the model moves from an adjacency matrix basis of centrality toward a basis that attends to the behaviors of users within the structure of the webgraph.

\section{Discussion}

This paper develops a generalization of the PageRank approach. The assumption of a scalar dampening coefficient is relaxed to allow for heterogeneous values. A visitation count approach may be employed to apply the more general model. The construct $\bf{W}$, which is conventionally based on the binary hyperlink structure of the webgraph, need not be so restricted. Further note that the necessary information to implement this approach is node-local, i.e., based on the number of visits to a node and the node's proportionate allocations of these visits to other nodes. A node's allocation weights ${w_{ij}}, \; (j= 1,2,..., n)$  may be selectively monitored for stability and updated.

\bibliography{ArticleReferences}

\end{document}